\documentstyle[preprint,aps,eqsecnum]{revtex} 
\input{epsf}
\def\be{\begin{equation}} 
\def\ee{\end{equation}}
\def\bea{\begin{eqnarray}} 
\def\eea{\end{eqnarray}}
\def\line{\hbox to \hsize}    
\def\frac #1#2{{#1\over #2}}

\def\det{{\rm det\,}}

\def \a{{\bf a}}
\def \b{{\bf b}}
\def \c{{\bf c}} 

\def \k{{\bf k}}

\def \v{{\bf v}}
\def \x{{\bf x}}
\def \hx{{\hat {\bf x}}}
\def \hy{{\hat {\bf y}}}
\def \hz{{\hat {\bf z}}}

\def\1{\mbox{\bf 1}}

\begin{document}
\draft 

\title{WAVE EQUATION FOR SOUND IN FLUIDS    
WITH VORTICITY}
\author{SANTIAGO ESTEBAN PEREZ BERGLIAFFA}
\address{Centro Brasileiro de Pesquisas F\'{\i}sicas\\
Rua Dr.~Xavier Sigaud 150, 22290-180 Urca\\
Rio de Janeiro, RJ-Brazil\\
E-mail: sepb@cbpf.br}
\author{KATRINA HIBBERD}
\address{Departamento de F{\'{\i}}sica Te{\'o}rica\\ 
     Universidad de Zaragoza\\ 
     50009 ZARAGOZA, 
     Spain \\
E-mail: keh@posta.unizar.es}
\author{ MICHAEL STONE}
\address{University of Illinois, Department of Physics\\
1110 W.~Green St., Urbana, IL 61801, USA\\E-mail: m-stone5@uiuc.edu}   
\author{ MATT VISSER} 
\address{Washington University, Physics Department\\
 Saint Louis, Missouri 63130-4899, USA\\E-mail: visser@kiwi.wustl.edu} 
 
\maketitle

\newpage
\begin{abstract}

We use Clebsch potentials and an action principle to derive 
a closed system of gauge invariant equations for sound  
superposed on a general background flow. Our system reduces
to  the Unruh (1981)  and   Pierce (1990) wave equations
when the flow is irrotational, or slowly varying.  We
illustrate our formalism by applying it to waves
propagating in a uniformly rotating fluid where the sound
modes hybridize with  inertial waves.

\end{abstract} 


\newpage 

\section{Introduction}

Studies of sound in a stationary fluid usually start from
from the wave equation whose derivation appears in all
elementary texts\cite{pierce81}. When the fluid  is moving, however,
finding an appropriate generalization of the wave equation 
is not always possible.  An exception is the special
circumstance  that the equation of state is barotropic and
the background flow irrotational, but not necessarily
steady.  In this case  a particularly attractive  equation
was derived by Unruh\cite{unruh81,unruh95}.  Unruh's  
equation was  later rediscovered and further popularized by
Visser\cite{visser98,visser99}.  It   coincides with the
equation obeyed by a relativistic scalar  field propagating
in  curved space-time.  The space-time geometry is 
governed by the  {\it acoustic metric} which depends on the
background flow velocity and on the local fluid density and
speed of sound.

The curved space-time  interpretation of the wave equation
is rather more than a mathematical curiosity. As well
providing  an attractive analogy with some aspects of
general relativity\cite{rio-conference}, one can use the
geometric formalism for ray tracing, and to produce  a
straightforward and systematic derivation of various
conservation laws associated with acoustic energy and
momentum\cite{stone00b,stone01}.  

Unfortunately most fluid motions  occurring in nature are
not irrotational. It is therefore desirable to explore the
possibility of extending the acoustic metric   equation to
a wider class of flows. A valuable step in this direction
was taken by Pierce\cite{pierce90}  who, without assuming  
that the background flow was irrotational,  derived an
equation which is in appearance equivalent  to the acoustic
metric  equation.  He did, however,  assume that the
background flow varied slowly  over  the  length and time
scale of the  sound wave. As a result of his
approximations,  his dependent variable is not quite  the
velocity potential appearing in the acoustic metric 
equation.  It  gives the fluctuating velocity only up to a 
correction whose magnitude depends on the space and time
inhomogeneities of the background flow. Pierce did not
attempt to characterize the correction beyond estimating
its size and showing that it was small in most regimes of
interest in acoustics. He did show, however, that in
special cases his equation reduces to  known exact wave
equations. In particular, for steady irrotational flow  it
reduces to Blokhintsev's equation\cite{blokhintsev45,blokhintzev46},
which is a special case of Unruh's. This and other features
led  Pierce to conjecture that his  equation is of wider
applicability  than his  derivation suggests.       The
purpose of the present paper is to show that this
conjecture is correct. 

We use Clebsch potentials and an action principle to derive the equations of
motion for small perturbations about a general barotropic
flow. Our  principal result is an exact and  concise
expression for the small correction to potential flow, and
a simple  equation of motion obeyed by it. This, coupled
with the Pierce equation, provides a closed system for wave
propagation in a general inhomogeneous and unsteady
background flow. The condition that the correction to
potential flow be ignorable, and thus the acoustic metric  
equation accurate, is that the frequency of the sound  be
appreciably higher than the local vorticity. There is {\it
no\/}  requirement that the spatial inhomogeneity be small.
Even if the frequency  condition is violated, we can
still  study the wave motion, but with a more
complicated system of partial differential equations.

The paper is organized as follows: In section
\ref{SEC:euler} we review the action principle for the
Clebsch  formulation of barotropic  fluid mechanics. In
section \ref{SEC:gauge} we discuss the infinite family of
conserved quantities that generate global gauge
transformations on the Clebsch potentials. These are needed
in section \ref{SEC:invariant} where we consider first
order perturbations to a background flow, and identify two
gauge invariant combinations of the potentials which have
physical significance. Sections \ref{SEC:wave} and
\ref{SEC:displacement} derive a   closed system of 
equations for  these combinations. Finally  section
\ref{SEC:illustration} illustrates our formalism by applying it
to waves in a uniformly rotating fluid.

\section{Euler Equation}
\label{SEC:euler}

In this section we will review the Clebsch potential
approach to fluid dynamics. The Clebsch formalism  has the
advantage that the equations of motion may be derived from
an action principle\cite{bateman29}, and with an action principle 
conservation laws  are related to symmetries by Noether's
theorem.

We begin with 
\be
S= \int d^4x \left\{ -\frac 12 \rho\v^2 - \phi(\dot
\rho+\nabla\cdot (\rho \v))+ \rho\beta (\dot \gamma +(\v\cdot
\nabla) \gamma)+u(\rho)\right\}.
\label{EQ:action}
\ee
Here $\rho$ is the fluid mass-density, $\v$ the velocity, and
$u(\rho)$  the internal energy density.    
This is the customary  expression  giving rise to  irrotational
fluid dynamics (see, for example \cite{stone00b})    ---
but with an additional term containing new fields: $\beta$
and $\gamma$.    The variable $\beta$ may be thought of as
a Lagrange multiplier enforcing the {\it Lin
constraint\/}\cite{lin63} that there be a label  ($\gamma$)
painted on the particles permitting  us  to distinguish 
one from another\footnote{Lin originally employed {\it
three\/} Lagrange multipliers $\beta_{1,2,3}$ leading to 
the conservation of three Lagrange co-ordinates,
$\gamma_{1,2,3}$,   which served to  label the material
particles uniquely. As shown by Seligar and
Whitham\cite{seligar68}, only one is really necessary.}.

Requiring that $S$ be stationary when we vary  $\v$ gives
\be
-\rho\v+\rho\nabla \phi +\rho\beta\nabla\gamma=0,
\ee
or
\be
\v= \nabla\phi+\beta\nabla\gamma.
\label{EQ:clebsch}
\ee
This  is the Clebsch representation\cite{clebsch1859,lamb32} of the
velocity field. It allows for flows with non-zero vorticity
\be
\omega =\nabla \times \v = \nabla \beta\times \nabla \gamma.
\ee

We use (\ref{EQ:clebsch})  to  algebraically eliminate  
the $\v$ in $S$ in favour  of the Clebsch potentials $\phi$,
$\beta$, $\gamma$. This leads to  a new action
\cite{bateman29} 
\be
S_{\rm new} = \int d^4x \left\{\frac 12\rho
(\nabla\phi+\beta\nabla\gamma)^2 +\rho(\dot \phi +\beta\dot
\gamma) +u(\rho)\right\}.
\label{EQ:newaction}
\ee

Varying the remaining variables in (\ref{EQ:newaction}) gives the equations of
motion
\bea
\delta \phi &:&\quad  \dot \rho +\nabla\cdot (\rho\v)=0,\nonumber\\
\delta \beta &:&\quad  \rho(\dot\gamma+(\v\cdot\nabla)
\gamma)=0\quad\Rightarrow\quad  \dot\gamma+(\v\cdot\nabla)
\gamma=0,\nonumber\\
\delta \gamma &:&\quad  \partial_t(\rho\beta)
+\nabla(\v\rho\beta)=0 \quad \Rightarrow \quad \dot \beta
+(\v\cdot\nabla)\beta=0,\nonumber\\
\delta \rho &:&\quad  \frac 12 v^2 +\dot\phi +\beta\dot\gamma
+\mu=0,  
\label{EQ:clebscheofm}
\eea
where, in the last line, $
\mu = {d u}/{d\rho}
$
is the specific enthalpy.
We see that the values of both $\beta$ and $\gamma$ are advected with the
motion.

We now show that these equations imply  the Euler
equation for the fluid. 
Apply   $\nabla_i$ to  the last line in
(\ref{EQ:clebscheofm}) and add and subtract 
$\dot\beta\nabla_i\gamma$,  so that the second term is the
time derivative of the velocity
\be
 v_k \nabla_i v_k +\partial_t(\nabla_i \phi
 +\beta\nabla_i \gamma) - \dot \beta\nabla_i \gamma
+\dot\gamma \nabla_i\beta = -\nabla_i\mu.
\ee  
In other words
\be
\partial_t v_i + v_k \nabla_i v_k - \dot \beta\nabla_i \gamma
+\dot\gamma \nabla_i\beta = -\nabla_i\mu.
\label{EQ:clebsch2}
\ee
The second, third, and fourth terms on the left hand side now need to be
taken care of.
Write
\bea
v_k \nabla_i v_k &=& v_k\nabla_k v_i +v_k (\nabla_i v_k
-\nabla_kv_i),\nonumber\\
&=& (\v\cdot \nabla) v_i +v_k(\nabla_i\beta\,\nabla_k\gamma
-\nabla_k\beta\,\nabla_i\gamma),\nonumber\\
&=& (\v\cdot \nabla) v_i - \dot\gamma \,\nabla_i \beta + \dot
\beta \,\nabla_i \gamma,
\label{EQ:this}
\eea 
where, in the last line, we have used the convective
constancy of $\beta$, $\gamma$.

Inserting  (\ref{EQ:this})  into (\ref{EQ:clebsch2}) we find
\be
\dot \v +(\v\cdot\nabla) \v= -\nabla \mu,
\ee
which is Euler's equation.

\section{Gauge Transformations}
\label{SEC:gauge}

In three dimensions any
vector field may be locally represented 
in the form\cite{clebsch1859,lamb32}  
\be
\v = \nabla \phi+ \beta\nabla \gamma.
\ee 
Given a velocity field $\v$, however, the potentials 
$\phi$, $\beta$ and $\gamma$, are not uniquely determined.
This indeterminacy  is usually  described as being  due
to a gauge invariance, but it is more analogous to the
residual freedom to make {\it time independent\/} gauge
transformations that survives  after we fix the  $A_0=0$
gauge  in Maxwell electrodynamics.  As in that example,
once we have a made a choice of  the potentials,  $\phi$,
$\beta$, and $\gamma$, at any particular time, their
subsequent evolution is uniquely determined by the
equations of motion (\ref{EQ:clebscheofm}).

We can relate the gauge invariance to conservation
laws. From
\bea
\dot\rho +\nabla\cdot \rho \v&=&0,\nonumber\\
\dot\beta +(\v\cdot \nabla)\beta &=&0,\nonumber\\
\dot\gamma +(\v\cdot \nabla) \gamma &=&0,
\eea
we deduce that 
\be
F=\int \rho {\cal F}(\beta, \gamma)d^3x 
\ee
is independent of time. Here ${\cal F}$ is an arbitrary
function of the variables $\beta$ and 
$\gamma$ with position independent coefficients.

Now  any action that contains only first order time
derivatives   defines a 
Poisson bracket  and canonical structure. 
For two functionals $F_{1,2}$ of the fields $\rho$,
$\phi$, $\beta$,  $\gamma$ at time $t$ we  
define the Poisson bracket $\{F_1,F_2\}$ as
\be
\{F_1,F_2\}= \left.\frac {d F_2}{dt}\right|_{F_1},
\ee
where the subscript, $F_1$, on the derivative indicates that 
time evolution of the variables $\rho$, $\phi$,
$\beta$, $\gamma$ is derived by varying the action
\be
S[F_1]= \int \rho(\dot \phi +\beta\dot
\gamma) \,d^4x -\int F_1(\rho, \phi, \beta, \gamma)dt.
\ee
Such a Poisson bracket automatically satisfies  all the usual properties, 
including skew symmetry 
and the  Jacobi identity
\be
\{F_1,\{F_2,F_3\}\}+
\{F_2,\{F_3,F_1\}\}+\{F_3,\{F_1,F_2\}\}=0.
\ee

In the present case the  bracket becomes 
\be
\{F_1,F_2\} = \int d^3x\left(
\frac 1\rho \frac{\delta F_1}{\delta \beta(\x)}
\frac{\delta F_2}{\delta \gamma(\x)} - \frac{\delta
F_1}{\delta \phi(\x)}
\frac{\delta F_2}{\delta \rho(\x)}- \frac \beta\rho \frac{\delta
F_1}{\delta \phi(\x)}
\frac{\delta F_2}{\delta \beta(\x)} - (F_1
\leftrightarrow F_2)\right),
\ee
and   
$(\rho$, $\phi)$, and
$(\rho\beta$, $\gamma)$   
constitute  two canonically conjugate pairs, {\it i.e.\/} 
\bea
\{\rho(\x),\phi(\x')\}=\delta^3(\x-\x'),\nonumber\\
\{\rho\beta(\x),\gamma(\x')\}=\delta^3(\x-\x').
\eea

We now consider  the conserved charge $F$ 
as the generator of an infinitesimal symmetry
by setting
\be 
\delta\phi = \{F, \phi\}= {\cal F } - \beta  \frac{\partial {\cal
F}}{\partial\beta}.
\ee
Similarly
\bea
\delta \beta &=& - \frac{\partial {\cal
F}}{\partial\gamma},\nonumber\\
\delta \gamma &=& \frac{\partial {\cal 
F}}{\partial\beta}.
\eea
The field $\rho$ is  unaltered. This is  because $F$ does not contain $\phi$.

These variations   generate  an infinite
dimensional   {\it global}  (rigid) symmetry group. It is a global
symmetry   because the parameters in ${\cal F}$ are 
required  to be independent of $\x$ and $t$. The
transformations   are  the extension to Clebsch potentials
of the   global $U(1)$ phase symmetry $\phi\to \phi+ const.$ 
which appears in  potential  flow, $\v=\nabla \phi$, 
where it is generated by the conserved
charge $Q=\int\rho \,d^3x$.

The symmetry transformations leave   the Hamiltonian 
\be
H =\int \left\{\frac 12\rho
(\nabla\phi+\beta\nabla\gamma)^2  +u(\rho)\right\}d^3x
\ee
invariant because $\{F, H\} = - \{H, F\} = \frac
{dF}{dt}=0$. In addition to Poisson-commuting with the hamiltonian, the
conserved charge $F$ generates 
variations  that
preserve $\v$ itself:
\bea
\delta \v &=& \nabla \delta\phi + \delta \beta \nabla\gamma
+ \beta\nabla\delta \gamma,\nonumber\\ 
&=& \nabla\left({\cal F}-\beta \frac{\partial {\cal
F}}{\partial\gamma}\right) -\frac{\partial {\cal
F}}{\partial\gamma}\nabla \gamma + \beta \nabla\left(\frac{\partial {\cal
F}}{\partial\beta}\right),\nonumber\\
 &=&0.
\eea
They also preserve the kinetic term:
\bea
&\delta(\rho(\dot \phi&+\beta \dot \gamma))\nonumber\\
\quad &=& \rho\left\{ \frac{\partial {\cal
F}}{\partial\beta}\dot \beta + \frac{\partial {\cal
F}}{\partial\gamma}\dot\gamma + \left(\frac{\partial {\cal
F}}{\partial t}\right)_{\beta,\gamma} - \dot \beta \frac{\partial {\cal
F}}{\partial\beta}- \beta\left(\frac{\partial^2 {\cal
F}}{\partial\beta^2}\dot\beta + \frac{\partial^2 {\cal
F}}{\partial\beta \partial \gamma}\dot\gamma +  \frac{\partial^2 {\cal
F}}{\partial t \partial \beta}\right)\right.\nonumber\\
 &&\quad -\left.\frac{\partial {\cal
F}}{\partial\gamma}\dot\gamma +\beta \left(\frac{\partial^2 {\cal
F}}{\partial\beta^2}\dot\beta +  \frac{\partial^2 {\cal
F}}{\partial \beta \partial \gamma}\dot\gamma +\frac{\partial^2 {\cal
F}}{\partial \beta \partial t}\right)\right\}\nonumber\\
&=& \rho \left(\frac{\partial {\cal
F}}{\partial t}\right)_{\beta\gamma},
\eea
which vanishes provided $F$ does not explicitly depend on time.

It is easy to show that 
the   symmetry  group  is  that of  orientation and area preserving
diffeomorphisms of the 2-plane. It is equivalently
the group of non-linear canonical
transformations on a two-dimensional phase space with 
Darboux co-ordinates  $\beta$, $\gamma$. Because of this
we  can  obtain the finite form of the
transformations --- as well as confirming  that that they
exhaust all  transformations that preserve $\v$ ---  by
exploiting the familiar generating function methods from
classical mechanics\cite{goldstein}. 

Suppose that 
\be
d\tilde \phi + \tilde\beta d \tilde \gamma = d\phi
+\beta d\gamma,
\ee
Then
\be
d(\tilde \phi -\phi) = \beta d\gamma - \tilde \beta d \tilde
\gamma,
\ee
and there must exist  a $W(\gamma,\tilde \gamma)$, the
{\it generating function\/,} 
such that
\be
\tilde \phi -\phi=W,\qquad \frac{\partial W}{\partial
\gamma} =\beta,\qquad \frac{\partial W}{\partial
\tilde \gamma} =\tilde \beta.
\ee
Conversely, given a generating function, we can obtain a
finite canonical transformation.

To make contact with the infinitesimal transformations we
considered earlier, we let
\bea
\tilde \beta =\beta + \dot \beta \Delta t,\nonumber\\
\tilde \gamma =\gamma + \dot \gamma \Delta t,
\eea
where ``$t$'' is a notional time parameterizing the change.

Thus
\bea
dW&=& \beta d\gamma - (\beta +  \dot \beta \Delta t)
           d( \gamma + \dot \gamma \Delta t),\nonumber\\
  &=& -\Delta t ( \dot \beta d\gamma +\beta d \dot \gamma).
\eea
Similarly let $W=U\Delta t$, so that
\be
dU= -\dot \beta d \gamma - \beta d\dot \gamma,
\ee
or, making a Legendre transformation $F=U+\beta\dot\gamma$,
\be
d(U+\beta\dot\gamma) = -\dot \beta d\gamma +\dot\gamma d\beta
= dF(\beta,\gamma).         
\ee    

In other words
\be 
\dot \beta = -\frac{\partial F}{\partial \gamma},\qquad \dot
\gamma= \frac{\partial F}{\partial \beta},
\ee
leading to
\bea
\tilde \phi &=& \phi + W \Delta t = \phi + \left(F-
\beta\frac {\partial F}{\partial\beta}\right)\Delta t,\nonumber\\
\tilde \beta &=&\beta -  \frac {\partial
F}{\partial\gamma}\Delta t,\nonumber\\
\tilde \gamma &=&\gamma +  \frac {\partial
F}{\partial\beta}\Delta t,
\eea
as before.

\section{Gauge-invariant Fluctuations}
\label{SEC:invariant}

We want to study  the evolution of small fluctuations $(\rho_1,
\phi_1,\beta_1, \gamma_1)$, superposed on  a background flow
$(\rho_0,\phi_0, \beta_0,\gamma_0) $. We will not assume
that this background  flow is steady, only  that it
satisfies the equations of motion. 

We expand the action  $S_{\rm new}$ out to quadratic order in
the fluctuations
\be
S_{\rm new}=S_0+S_1+S_2+\cdots .
\ee
The action $S_1$, containing terms linear in the
fluctuations,  vanishes because of
our assumption that the zeroth order variables obey the
equation of motion. The term quadratic in the fluctuations 
is
\be
S_2= \int d^4x\left\{\frac 12 \rho_0 \v_1^2 +\rho_1
\v_0\cdot \v_1 +\rho_1(\dot \phi_1+
\beta_0\dot\gamma_1+\beta_1\dot \gamma_0) +\rho_0\beta_1
\dot \gamma_1 +\frac 12 \frac{c^2}{\rho_0} \rho_1^2\right\}, 
\label{EQ:S2}
\ee
where $\v_1$ is shorthand  for $\nabla\phi_1 +\beta_1
\nabla\gamma_0 + \beta_0 \nabla\gamma_1$, and
\be
{c^2}= \rho_0\,\frac{d^2 u}{d\,\rho^2},
\ee
is the square of the local speed of sound.
From $S_2$ we can deduce the equations of motion for the
first-order fluctuating quantities. These equations are not
easy to work with, however. Because they are
advected with the flow, the potentials
$\beta_0$ and $\gamma_0$ which  appear as coefficients in the
equations  will generally be time dependent --- even  if the
background flow is steady. Furthermore, the overall gauge
ambiguity obscures any physical interpretation.   
It is therefore fruitful to seek combinations of the
potentials that are gauge invariant and can  be expressed
in terms of physical quantities. For example the  first
order velocity field, 
\be
\v_1= \nabla \phi_1 + \beta_0 \nabla \gamma_1 + \beta_1
\nabla \gamma_0,
\ee
is gauge invariant because $\v$ is.

By varying $\rho_1$ in (\ref{EQ:S2}) we find  
\be
\rho_1 = -\frac{\rho_0}{c^2}\left( \dot \phi_1 +\v_0\cdot
\nabla \phi_1 +\beta_0( \dot\gamma_1 +\v_0\cdot
\nabla \gamma_1) +\beta_1 (\dot\gamma_0 +\v_0\cdot
\nabla \gamma_0)\right).
\ee
Since  
\bea
\dot\beta_0 +\v_0\cdot
\nabla \beta_0&=&0,\nonumber\\
\dot\gamma_0 +\v_0\cdot
\nabla \gamma_0&=&0,
\eea
we can write this as 
\be
\rho_1 = -\frac{\rho_0}{c^2}\frac {d\psi_1}{dt},
\label{EQ:dpsi}
\ee
where 
\be
\psi_1 =\phi_1 +\beta_0 \gamma_1, 
\ee
and
\be
\frac {d}{dt}= \frac {\partial}{\partial t} + \v_0\cdot
\nabla
\ee
is the convective derivative.

The density fluctuation $\rho_1$, being a physical variable, is gauge
invariant. Consequently  equation  (\ref{EQ:dpsi}) suggests that
the combination $\psi_1$ is itself  gauge invariant. This is
easily confirmed:
\bea
\delta(\phi_1 +\beta_0 \gamma_1) &=& \left\{ \frac{\partial {\cal
F}}{\partial\beta} \beta_1 + \frac{\partial {\cal
F}}{\partial\gamma}\gamma_1  -  \beta_1 \frac{\partial {\cal
F}}{\partial\beta}- \beta_0\left(\frac{\partial^2 {\cal
F}}{\partial\beta^2}\beta_1 + \frac{\partial^2 {\cal
F}}{\partial\beta \partial \gamma}\gamma_1 \right)\right.\nonumber\\
 &&\quad -\left.\frac{\partial {\cal
F}}{\partial\gamma}\gamma_1 +\beta_0 \left(\frac{\partial^2 {\cal
F}}{\partial\beta^2}\beta_1 +  \frac{\partial^2 {\cal
F}}{\partial \beta \partial \gamma}\gamma_1 \right)\right\}\nonumber\\
&=& 0.
\eea

We can use $\psi_1$ to write 
\be
\v_1= \nabla \psi_1 +\xi_1,
\label{EQ:key}
\ee
where 
\be
\xi_1= \beta_1 \nabla\gamma_0 - \gamma_1
\nabla \beta_0.
\label{EQ:psidef1}
\ee
This is  a decomposition of the first-order velocity fluctuation
into two gauge invariant parts. Because sound in a fluid is a scalar
excitation, it is natural to identify the scalar field
$\psi_1$ with  the acoustic degree of freedom, and $\xi_1$,
the correction to potential flow induced  by angular
momentum conservation, with a  partial
hybridization of the sound with other  modes.     

Although the vector field $\xi_1$ has three components, it only
represents two degrees of freedom. This is  because  
\be
\xi_1 \cdot \omega_0 = (\beta_1 \nabla\gamma_0 - \gamma_1
\nabla \beta_0)\cdot (\nabla\beta_0 \times \nabla
\gamma_0)\equiv 0.
\ee

Since $\xi_1$ is gauge invariant, it should be possible
to write  it in terms of physical variables. In 
section \ref{SEC:displacement} we will show that it is equal to $\x_1\times
\omega_0$ where $\x_1$ is the particle displacement caused by
the disturbance.

\section{Wave Equation}
\label{SEC:wave}

The first-order continuity equation
\be
\frac{\partial \rho_1}{\partial t} + \v_0\cdot \nabla \rho_1
+  \rho_1\nabla\cdot \v_0 +  \nabla\cdot \rho_0\v_1=0,
\ee
together with the zeroth order continuity equation
\be
\frac{\partial \rho_0}{\partial t} +  \nabla\cdot 
\rho_0\v_0=0,
\label{EQ:zeroth-continuity}
\ee
the equation for $\rho_1$,
\be
\rho_1= -\frac {\rho_0}{c^2}\frac {d\psi_1}{dt}, 
\ee
and the  decomposition
$
\v_1= \nabla\psi_1 +\xi_1,
$
may be combined to give
\be
\frac {d}{dt} \left(\frac 1{c^2}\frac {d}{dt}\psi_1\right)=\frac 1
{\rho_0} \nabla \left(\rho_0(\nabla \psi_1+\xi_1)\right).
\label{EQ:pierceexact}
\ee

If we ignore the  $\xi_1$, (\ref{EQ:pierceexact}) is
Pierce's approximate wave equation
\be
\left(\frac{\partial}{\partial t}+\v_0\cdot\nabla\right)\frac
1{c^2}\left(\frac{\partial}{\partial t}+\v_0\cdot\nabla\right)\psi_1=
\frac 1 {\rho_0}\nabla(\rho_0\nabla\psi_1).
\label{EQ:pierce}
\ee
By using (\ref{EQ:zeroth-continuity}) again, this can be rewritten as  
\be
\left(\frac{\partial}{\partial t}+\nabla\cdot\v_0\right)\frac
{\rho_0}{c^2}\left(\frac{\partial}{\partial t}+\v_0\cdot\nabla\right)\psi_1=
\nabla(\rho_0\nabla\psi_1),
\label{EQ:unruheq}
\ee
where each $\nabla$ is acting on everything to its right.

Although (\ref{EQ:pierce}) may seem more natural, 
the form (\ref{EQ:unruheq}) 
has the advantage that it can be written as\footnote{We use the convention 
that greek letters 
run over four space-time 
indices  $0,1,2,3$ with $0\equiv t$, while roman indices refer 
to the three space
components.}
\be
\frac 1{\sqrt{-g}} \partial_\mu {\sqrt{-g}}g^{\mu\nu}\partial_\nu
\psi_1=0,
\label{EQ:scalar_eq}
\ee
where 
\be
\sqrt{-g}g^{\mu\nu} = 
\frac {\rho_0}{c^2}
\left(\matrix{ 1, &\v_0^T \cr
             \v_0,& \v_0\v_0^T - c^2{\bf 1}\cr}\right).
\label{EQ:unruh_metric_up}
\ee  
Equation  (\ref{EQ:scalar_eq})   has the same form as  that
of a scalar wave propagating  in a  gravitational field
with pseudo-Riemann (Lorentzian)  metric $g_{\mu\nu}$. We will refer to
$g_{\mu\nu}$ as the acoustic  metric. The idea of writing
the  sound wave equation in this way is due to Unruh
\cite{unruh81,unruh95}.  

As is customary in general relativity, the symbol $g$ denotes the
determinant of the covariant form of the metric,
$g_{\mu\nu}$, so $\det g^{\mu\nu}=g^{-1}$. Taking the
determinant of both sides of (\ref{EQ:unruh_metric_up}) thus
shows that the $4$-volume measure   $\sqrt{-g}$
is equal to $ \rho_0^2/c$. Knowing  this, we may then
invert the matrix $g^{\mu\nu}$ to find  
the covariant components of the metric
\be
g_{\mu\nu}= \frac {\rho_0}{c}
\left(\matrix{ c^2-v_0^2, & \v_0^T \cr
                     \v_0,& -{\bf 1}\cr}\right).
\label{EQ:unruh_metric_down}
\ee
The associated space-time interval is therefore
\be
ds^2= \frac {\rho_0} c \left\{c^2dt^2
-\delta_{ij}(dx^i-v_0^idt)(dx^j-v_0^jdt)\right\}.
\label{EQ:ADM}
\ee
In the geometric acoustics limit, sound propagates along the
null geodesics of this metric.

Metrics of the  form (\ref{EQ:ADM}), although without the overall
conformal factor ${\rho_0}/{c}$, appear in the 
Arnowitt-Deser-Misner (ADM) formalism of general
relativity\cite{ADM}.  There, $c$ and $-v_0^i$ are referred
to as  the {\it lapse function\/} and {\it shift vector\/} 
respectively. They serve to glue successive
three-dimensional time slices together to form a
four-dimensional space-time\cite{MTW}. In our present case,
provided  ${\rho_0}/{c}$ can be regarded as a
constant,  each $3$-space is ordinary flat ${\bf R^3}$
equipped with the rectangular cartesian metric
$g^{(space)}_{ij}=\delta_{ij}$ --- but the resultant
space-time is in general curved, the curvature depending on
the degree of inhomogeneity of the mean flow $\v_0$.

This formalism is very pretty, but (\ref{EQ:pierce}) is
exact only when the background flow is potential.
Equation (\ref{EQ:pierceexact}), on the other hand,  
is valid for a general barotropic flow ---  but to
be of use it must be complemented by an equation
determining the time evolution of $\xi_1$. We now derive
such an  equation.

We start with the observation that, since  $\beta$, $\gamma$ are 
convectively conserved, we have 
\be
\frac{\partial \beta_0}{\partial t} + (\v_0\cdot \nabla)
\beta_0 =0,
\label{EQ:beta0t}
\ee
and
\be
\frac{\partial \beta_1}{\partial t} + (\v_0\cdot \nabla)
\beta_1 + (\v_1 \cdot \nabla) \beta_0=0.
\ee
Taking the gradient of (\ref{EQ:beta0t}) gives
\be
\left(\frac {\partial}{\partial t} + \v_0\cdot \nabla\right)
\nabla_i \beta_0 = - (\nabla_i v_{0j}) \nabla_j
\beta_0. 
\ee

Thus, using the definition (\ref{EQ:psidef1}),
\bea
\left(\frac {\partial}{\partial t} + \v_0\cdot \nabla\right)
\xi_{1i} &=& -[(\v_1 \cdot \nabla) \beta_0]\nabla_i \gamma_0 
           + [(\v_1 \cdot \nabla) \gamma_0]\nabla_i
	   \beta_0\nonumber\\
	   &&\quad -\beta_1 (\nabla_i v_{0j}) \nabla_j
	   \gamma_0  +\gamma_1 ( \nabla_i v_{0j}) \nabla_j
	   \beta_0\nonumber\\
      &=& -v_{1j}(\nabla_j \beta_0 \nabla_i \gamma_0 - 	   
	     \nabla_j \gamma_0 \nabla_i \beta_0) -
	     (\nabla_i v_{0j})\xi_{1j}\nonumber\\
      &=& -v_{1j}(\nabla_j v_{0i} - \nabla_i v_{0j}) 
           - (\nabla_i v_{0j})\xi_{1j}\nonumber\\
      &=& (-\nabla_j\psi_1 -\xi_{1j})(\nabla_j v_{0i} -
      \nabla_i v_{0j})	   	     
           -(\nabla_i v_{0j})\xi_{1j},
\label{EQ:xieqofmot1}
\eea
which can be written as
\be
\frac {d \xi_1}{dt} = \nabla \psi_1 \times \omega_0 - (\xi_1\cdot
\nabla)\v_0.
\label{EQ:xidot}
\ee

The two equations 
\be
\frac {d}{dt} \left(\frac 1{c^2}\frac {d}{dt}\psi_1\right)=\frac 1
{\rho_0} \nabla \left(\rho_0(\nabla \psi_1+\xi_1)\right).
\ee
and
\be
\frac {d \xi_1}{dt} = \nabla \psi_1 \times \omega_0 - (\xi_1\cdot
\nabla)\v_0,
\ee
form a closed system of equations, containing only
gauge-invariant quantities,  describing the first-order
fluctuations about  the background mean flow.

\section{Displacement field}
\label{SEC:displacement}

It is not yet clear that, under most circumstances 
of interest in acoustics, the quantity  $\xi_1$ 
is a small correction to $\nabla\psi_1$.  
It becomes so, however, once  we establish the
result 
\be
\xi_1 = \x_1 \times \omega_0
\label{EQ:xieq}
\ee
where $\x_1$ is the displacement of a material particle 
due to the sound wave.
Given  (\ref{EQ:xieq}), we see that the order of magnitude
of $\xi_1$ is that of the  product of the displacement 
amplitude with the background flow rotation frequency. The
fluctuating velocity associated with the acoustic
field is, on the other hand, of the order of  the
displacement amplitude times the frequency, $\Omega$, of
the sound wave.  Thus\footnote{Our  argument tacitly
assumes that $\x_1$ remains small and oscillating. This is
certainly what we expect for a sound wave, but, in the
absence of viscous damping,  many flows with vorticity will
be unstable to the onset of turbulence, and if the sound
triggers such an instability $\x_1$ will grow without
bound.}  $\xi_1$ is smaller than $\nabla\psi_1$ by a factor
of $|\omega_0|/\Omega$.           

To establish (\ref{EQ:xieq}) we first define  $\x_1(\x,t)$
to be  the eulerian displacement field (i.e the material
point which in the unperturbed reference flow was at time
$t$  located at $\x$ is, as a result of the perturbation,
now to be found  at $\x+\x_1$). We  remember that the
numerical values of the potentials   $\beta$, $\gamma$, 
are painted  on the  material particles, and so move with
the flow under both time evolution and the creation of an 
initial perturbation by  means of an 
external potential body force. Interpreting
this statement mathematically leads to  
\bea
\x_1 \cdot\nabla \beta_0 +\beta_1&=&0,\nonumber\\
\x_1 \cdot\nabla \gamma_0 +\gamma_1&=&0.
\eea
From this we may write
\bea
\beta_1 \nabla\gamma_0 - \gamma_1
\nabla \beta_0 &=& (\x_1 \cdot\nabla \gamma_0) \nabla\beta_0 -
(\x_1 \cdot\nabla \beta_0 )\nabla\gamma_0,\nonumber\\
&=& \x_1 \times (\nabla\beta_0\times\nabla
\gamma_0),\nonumber\\
&=&  \x_1 \times \omega_0.
\label{EQ:displacement}
\eea 
 
We can use  (\ref{EQ:xieq}) to rederive the equation of
motion for $\xi_1$ and so provide a derivation of  the wave
equation that is independent of the use of Clebsch
potentials. In their absence, though, the origin of 
the decomposition of the velocity field into the sum of
$\xi_1=\x_1\times \omega_0$  and the gradient of the pressure potential,
$\psi_1$, is a trifle obscure.  

To verify that (\ref{EQ:xieq}) leads to the equation of motion
(\ref{EQ:xieqofmot1}) 
for $\xi_1$  we must 
first establish a  connection between $\v_1$ and  the time
derivative of $\x_1$.  This requires us to describe  the
perturbation with  a little more formality.
Consider a family $\v(\x,t,\lambda)$ of adjacent solutions
of the full equations of motion. The velocity field
$\v(\x,t,0)$ is that of the unperturbed reference flow, and
increasing values of $\lambda$ correspond to flows evolving
from a one-parameter family of initial perturbations. 
By definition the  operations of time evolution and variation
of $\lambda$ commute. 

The
position, $\x(t,\lambda)$, of a material particle is given
by the solution to the differential equation
\be
\dot \x(t) = \v(\x(t,\lambda),t, \lambda)
\label{EQ:dotx}
\ee  
with suitable initial conditions.
Our  first order perturbed fields are, in this language, 
\be
\x_1 = \left.\frac {d\x}{d\lambda}\right|_{\lambda=0}, \qquad \v_1= 
\left.\frac{d\v}{d\lambda}\right|_{\lambda=0}.
\ee
Differentiating (\ref{EQ:dotx}) with respect to  $\lambda$,
and interpreting  the time derivative as a convective
derivative, gives
 \be 
\v_1 = \frac{\partial \x_1}{\partial t} + 
(\v_0\cdot\nabla)\x_1 - (\x_1\cdot \nabla) \v_0.
\label{EQ:vfromx1}
\ee


Now, starting from 
\be
\xi_1= \x_1\times \omega_0
\ee
and  the convective derivatives
\bea
\frac {d \x_1}{dt} &=& \v_1 + (\x_1\cdot \nabla) \v_0\\
\frac {d \omega_0}{dt} &=& (\omega_0\cdot \nabla) \v_0 -
(\nabla\cdot \v_0)\omega_0,
\eea
we may find an equation for the  time  evolution of $\xi_1$.
Using the fact the convective derivative is a derivation, we
find
\bea 
\frac {d \xi_1}{dt}&=& (\v_1 + (\x_1\cdot \nabla)
\v_0)\times\omega_0 + \x_1 \times((\omega_0\cdot \nabla) \v_0 -
(\nabla\cdot \v_0)\omega_0)\nonumber\\
&=& \v_1\times\omega_0 
+ (\nabla \cdot \x_1)(\v_0 \times \omega_0) 
+ (\nabla \cdot \omega_0)(\x_1 \times \v_0) 
+ (\nabla \cdot \v_0)(\omega_0 \times\x_1).
\eea
In the second line  the ordering of the symbols is meant only
to indicate how the indices are wired up. The $\nabla$ must
be  understood to act to the
right only on  the velocity field $\v_0$. 

We now use the vector identity 
\be
 (\x\cdot \a)(\b\times\c) 
+(\x\cdot \b)(\c\times\a) 
+(\x\cdot \c)(\a\times\b) = \x [\a\cdot(\b\times\c)]
\ee
with $\x$ replaced by $\nabla$ (still acting only on $\v_0$) to   find that 
\bea
\frac {d \xi_1}{dt}&=& \v_1\times\omega_0 + \nabla (\x_1\cdot
(\v_0\times \omega_0))\nonumber\\
&=& \v_1\times\omega_0 - \nabla (\v_0\cdot
(\x_1\times \omega_0))\nonumber\\
&=&\v_1\times\omega_0 - \nabla (\v_0\cdot
\xi_1)\nonumber\\
&=& (\nabla\psi_1)\times \omega_0 -(\xi_1\cdot \nabla)\v_0
\eea
which is the same as (\ref{EQ:xidot}).
(Again, in the first three lines,  $\nabla$ must be understood
to act only on $\v_0$,
even though it may be written to the left of other
variables.)
 
We can also check the consistency of the time evolution of
the first order vorticity.
From (\ref{EQ:displacement}) we find that 
\be
\omega_1 =\nabla\times(\x_1\times\omega_0),
\label{EQ:omegafromx}
\ee
so
\be
\frac{\partial\omega_1}{\partial t}
=\nabla\times \left(\frac{\partial \x_1}{\partial
t}\times\omega_0\right)+
\nabla\times\left(\x_1\times\frac{\partial \omega_0}{\partial
t}\right).
\label{EQ:dotomega1}
\ee 
It is not immediately obvious that (\ref{EQ:dotomega1}) is  
compatible with the equation 
\be
\frac{\partial\omega_1}{\partial t}
=  \nabla\times(\v_0 \times\omega_1)+
\nabla\times(\v_1 \times \omega_0),
\label{EQ:dotomega2}
\ee
which comes from applying $d/d\lambda$ to the 
vorticity evolution equation
\be
\frac{\partial \omega}{\partial t} = \nabla\times(\v\times
\omega).
\ee

The  right hand sides of (\ref{EQ:dotomega1}) and
(\ref{EQ:dotomega2}) are equal only if
\be
\frac{\partial\x_1}{\partial t} \times \omega_0 +
\x_1\times \frac{\partial \omega_0}{\partial t} 
- \v_0 \times \omega_1 -\v_1\times \omega_0,
\label{EQ:thing}
\ee 
is the gradient of something.
Now by using (\ref{EQ:vfromx1}), (\ref{EQ:omegafromx}) and
(\ref{EQ:dotomega2}), we can write  {(\ref{EQ:thing})  as
\bea
 \omega_0\times\left(\nabla\times(\x_1    \times\v_0)\right) &+& 
     \x_1\times\left(\nabla\times(\v_0    \times \omega_0)\right) 
  + \v_0 \times\left(\nabla\times(\omega_0\times \x_1)\right)\nonumber\\
 - (\omega_0 \times \x_1)     (\nabla\cdot \v_0) 
&-&  (\x_1   \times \v_0)     (\nabla\cdot \omega_0)
 -   (\v_0   \times \omega_0) (\nabla\cdot \x_1).
\label{EQ:thing2}
\eea
Here we have added in a term $(\x_1 \times \v_0) (\nabla\cdot
\omega_0)$, which is of course identically zero, in order to
preserve manifest cyclic symmetry of the terms.
  
Now for any three vector fields $\a$, $\b$, $\c$,
we may verify that 
\bea
\a \times (\nabla\times(\b\times \c)) +
\b \times (\nabla\times(\c\times \a)) &+&
\c \times (\nabla\times(\a\times \b))\nonumber\\
- (\a\times \b)(\nabla \cdot \c)
- (\b\times \c)(\nabla \cdot \a) 
&-& (\c\times \a)(\nabla \cdot \b)
\nonumber\\
= \nabla(\a\cdot(\b\times\c)),&&
\eea
where $\nabla$ is acting on everything to its right.
Applying this to (\ref{EQ:thing2}) shows that it {\it is\/}
a total derivative, and so the evolution
equations are consistent.

\section{Illustration}
\label{SEC:illustration}

As an illustration of the formalism consider   waves
propagating in the background flow
\be
\v_0= \frac{\omega_0}{2}\left(\matrix{-y\cr x\cr
0\cr}\right).  
\ee
This corresponds to the fluid rotating as a rigid body with
angular frequency $\omega_0/2$.
The perverse notation for the frequency arises  because we have
been using the symbol $\omega$ to denote  vorticity, and  $\nabla
\times \v_0= \omega_0 \,\hz$. 
 
To reduce notational clutter, in this section we will drop the 
suffix $1$ from the fields
$\psi$ and $\xi$. It should still be borne in mind  that they are
first order quantities.

\subsection{Sound/Inertial-Wave Hybridization}     

Our equations of motion are 
\bea
- \frac{d}{dt} \left(\frac 1 {c^2} \frac{d\psi}{dt}\right)
+\frac {1}{\rho_0} \nabla \rho_0 (\nabla \psi
+\xi)&=&0,\nonumber\\
\frac {d \xi}{dt} - (\nabla\psi\times \omega_0) +(\xi\cdot
\nabla) \v_0&=&0.
\eea
They need to be supplemented  with an initial condition that
sets $\xi\cdot\omega_0=0$. This orthogonality is then
preserved by the subsequent motion. We will ignore any
effects due to gradients in $\rho_0$ and $c^2$.

Take as an {\it ansatz \/}  a plane wave in the frame rotating with the fluid.
\bea
\xi &=& (\xi_{x'} \hx' +\xi_{y'} \hy')= (\Xi_{x'} \hx'
+\Xi_{y'} \hy')\,
e^{i(k_{x'} x'+k_{y'} y'+k_{z'} z'-\Omega
t)},\nonumber\\
\psi &=& \Psi \,e^{i(k_{x'} x'+k_{y'} y'+k_{z'} z'-\Omega t)}. 
\eea   
Here $\Xi_{x',y'}$ and $\Psi$ are constant amplitudes. The primed
unit vectors are  
\bea
\hx' &=& \hx \cos \left(\frac{\omega_0}{2}\right) t 
+ \hy\sin \left(\frac{\omega_0}{2}\right) t,
\nonumber\\
\hy'&=& -\hx \sin \left(\frac{\omega_0}{2}\right) t 
+ \hy\cos \left(\frac{\omega_0}{2}\right) t,
\nonumber\\
\hz'&=&\hz.
\eea
and the primed coordinates 
\bea
x' &=& x \cos \left(\frac{\omega_0}{2}\right) t 
+ y\sin \left(\frac{\omega_0}{2}\right) t,
\nonumber\\
y'&=& -x \sin \left(\frac{\omega_0}{2}\right) t 
+ y\cos \left(\frac{\omega_0}{2}\right) t,
\nonumber\\
z'&=&z.
\eea

The  convective derivatives on $\psi$ and on the
components of $\xi$ become
\bea
\frac {d\psi}{dt}&=& \left(\frac{\partial\psi}{\partial
t}\right)_{x',y'} =   -i\Omega \psi,\nonumber\\
\frac {d\xi_{x',y'}}{dt}&=& \left(\frac{\partial\xi_{x',y'}}{\partial
t}\right)_{x',y'} =   -i\Omega \xi_{x',y'}.
\eea
For  $\xi$ itself we need to take  note  of the time dependence 
of the the unit vectors $\hx'$, $\hy'$, so we
have
\bea
\frac {d\xi}{dt}&=& \left( \frac{d \xi_{x'}}{ dt}
-\left(\frac{\omega_0}{2}\right)
\xi_{y'}\right)  \hx' +
\left( \frac{d \xi_{y'}}{d t}
+\left(\frac{\omega_0}{2}\right)\xi_{x'}\right)  \hy'\nonumber\\
&=& (-i\Omega \xi_{x'} -\left(\frac{\omega_0}{2}\right)
\xi_{y'})\hx' +
(-i\Omega\xi_{y'} +\left(\frac{\omega_0}{2}\right) \xi_{x'})\hy'.
\eea

Also we need 
\be
(\xi\cdot\nabla)\v_0= -\xi_{y'} \left(\frac{\omega_0}{2}\right)
\hx' +\xi_{y'}
\left(\frac{\omega_0}{2}\right) \hy'.
\ee

The two off-diagonal $\omega_0/2$ terms add to get rid of
the $1/2$. The coupled equations therefore become
\be
\left(\matrix{ -i\Omega   & -\omega_0&-ik_{y'}\omega_0 \cr   
                +\omega_0& -i\Omega  &+ik_{x'}\omega_0 \cr
	       +ik_{x'}  & ik_{y'}  &\left(\frac{\Omega^2}{c^2}-k^2\right) \cr}
\right)	       
\left(\matrix {\Xi_{x'} \cr \Xi_{y'} \cr \Psi \cr}\right) =0.
\label{EQ:matrixeq}
\ee

For a solution to exist, the determinant of the matrix in
(\ref{EQ:matrixeq}) must vanish. This
gives the dispersion relation 
\be
(\omega_0^2-\Omega^2)\left(\frac{\Omega^2}{c^2}-|k|^2\right)+\omega_0^2
k_{x'}^2 
+ \omega_0^2 k_{y'}^2 =0,
\ee
which for fixed $\k$ is a quadratic equation for $\Omega^2$.

Some insight into this dispersion relation can be obtained
by letting $c^2\to \infty$. In this limit  the quadratic
reduces to
\be
|k|^2\Omega^2 -\omega_0^2 k_{z'}^2=0,
\ee
and so  gives
\be
\Omega^2= \frac{\omega_0^2 k_{z'}^2}{k_{x'}^2+k_{y'}^2+k_{z'}^2}.
\ee 
This is   the well-known dispersion relation for inertial
waves in an incompressible fluid\cite{lighthill}. For these modes  the
restoring force comes entirely from angular momentum conservation. 
They are  low frequency,   $\Omega^2\le \omega_0^2$,
oscillations and have a number of unusual features. 
In particular  the frequency is independent of the magnitude of
$\k$, so the group velocity is perpendicular to the phase
velocity --- {\it i.e.} parallel to the  wavecrests. At any
particular frequency the disturbance spreads out from its
source along a diabolic cone.

The second  root of the quadratic equation, 
$\Omega^2\approx c^2 k^2$ corresponds to  conventional
sound, and is lost to  infinity as $c^2$ becomes large.

Now let us consider general values of $c^2$.
From the eigenmode equation we can solve for $\xi$ in terms
of the amplitude of $\psi$ to get
\be
\left(\matrix {\Xi_{x'} \cr \Xi_{y'} \cr }\right)=
\frac{\omega_0}{\Omega^2-\omega_0^2}
\left(\matrix {-k_{y'}\Omega + ik_{x'}\omega_0 \cr k_{x'}\Omega
+ik_{y'}\omega_0  \cr }\right)\Psi.
\ee
This appears to be singular when $\Omega^2$ approaches $\omega_0^2$, but,
as we will see, this  occurs only near $k_{x'}=k_{y'}=0$ and 
the limit is smooth, the fluid rotating in circles in the
$x$-$y$ plane.  
   
From $\xi$  we can find the velocity field, $\v_1$, and hence, by
integration,  the
first order displacement field, $\x_1$, in the
frame rotating with the background fluid\footnote{If  
$
\x_1 = x_{1x'} \hx' +x_{1y'} \hy',
$
and 
$ 
\v_1 = v_{1x'} \hx' +v_{1y'} \hy',
$
then (\ref{EQ:vfromx1}) reduces to 
$
v_{1x',y'} =\left(\frac{\partial x_{1x',y'}}{\partial
t}\right)_{x',y'}.$}.
We therefore  find
\be
\x_1= \left(\frac {\Psi} {\Omega}\right) \left[
\left(\matrix {-k_{x'} \cr -k_{y'} \cr -k_{z'} \cr}\right)
+ \frac{ \omega_0}{\Omega^2-\omega_0^2}
\left(\matrix {-k_{x'}\omega_0 - ik_{y'}\Omega  \cr 
               -k_{y'}\omega_0 + ik_{x'}\Omega\cr 
	         0                              \cr}\right)
		  \right] e^{i(k_{x'} x'+k_{y'}y'+k_{z'} z'-\Omega t)}.
\ee

It is now straightforward to verify 
that we recover $\xi$ from $\xi = (\x_1\times
\omega_0)$. We also verify that the correction to  potential
flow is  $O(\omega_0/\Omega)$ when $\Omega\gg \omega_0$.

\subsection{Poincar{\'e} Waves}

If we restrict ourselves waves  with  $k_z=0$, 
then setting the  determinant  
to zero gives
\be
\frac {\Omega^2}{c^2} \left( \omega_0^2 + c^2 k^2
-\Omega^2\right)=0.
\ee
We therefore have two classes of modes: those with
zero frequency, and those with a gapped dispersion relation
\be
\Omega^2 =\omega^2_0 +c^2 (k_{x'}^2+k_{y'}^2).
\label{EQ:poincare'}
\ee
The former are $z$ independent
(Taylor-column\cite{taylor22,proudman16})  geostrophic flows where
pressure gradients are balanced against a Coriolis force.
The gapped modes are the Poincar{\'e}
waves\cite{pedlosky86}. 
 
We can obtain the Poincar{\'e} modes by considering the
 motion directly in the $x'$, $y'$  frame. The effect
of the frame rotation produces a  Coriolis force, and so 
the equation of motion is
\bea
\frac {\partial v_{1x'}}{\partial t} &=& \omega_0   v_{1y'}
-c^2\nabla_{x'}
\rho_1,\nonumber\\
\frac{\partial  v_{1y'}}{\partial t} &=& -\omega_0 v_{1x'}
-c^2 \nabla_{y'}
\rho_1.
\eea
To solve we need to combine this with   continuity equation
\be
\frac{\partial \rho_1}{\partial t} + \rho_0 \left(\nabla_{x'}v_{1x'} +
\nabla_{y'} v_{1y'}\right)=0.  
\ee

For waves travelling in the $\hx'$ direction we find   
\bea
v_{1x'}&=& A \cos(kx'-\Omega t),\nonumber\\
v_{1y'}&=& A \left(\frac {\omega_0} \Omega\right)
\sin(kx'-\Omega t),\nonumber\\
\rho_1 &=&A
\left(\frac{\rho_0k}{\Omega}\right)\cos(kx'-\Omega t),
\eea 
together with the dispersion relation (\ref{EQ:poincare'}). 


We also find the displacements of the particles to be
\bea
x_{1x'} &=&- A\left(\frac {1}{\Omega}\right)\sin(kx'-\Omega
t),\nonumber\\
x_{1y'} &=& A
\left(\frac{\omega_0}{\Omega^2}\right)\cos(kx'-\Omega t).
\eea
When  $\k=0$, we have $\Omega=\omega_0$, and  the particles move  in
circles in the $x$, $y$ plane. This limiting motion is
solenoidal and coincides with the $k_{x'}=k_{y'}=0$ limit of the
incompressible fluid inertial waves.

We may  now make contact with our $\psi$, $\xi$ formalism  by 
writing
\be
\rho_1= -\frac{\rho_0}{c^2} \frac {d
\psi}{dt} = \frac{\rho_0}{c^2} \left(\frac {\partial
\psi}{\partial t}\right)_{x',y'}
\ee
where
\be
\psi= A\left(\frac{c^2 k }{\Omega^2}\right) \sin(kx'-\Omega
t).
\ee

The relation $ \xi = \x_1 \times \omega_0$, which was
derived in the $x$, $y$ inertial frame,  continues to hold in the
rotating frame without modification. So we  have 
\be
\xi = \x_1 \times \omega_0 = \omega_{0z} (x_{1y'}  \hx' -
x_{1x'} \hy').
\ee
We  can combine our  expression for $\psi$ with this
to get   
\bea
v_{1x'} = \nabla_{x'} \psi + \xi_{x'} &=& A\left( \frac
{c^2k^2}{\Omega^2}+ \frac{ \omega_0^2}{\Omega^2}\right) 
\cos(kx'-\Omega t)\nonumber\\
v_{1y'} = \,\, 0\,\, + \xi_{y'} &=& A \left( \frac \omega \Omega\right)
\sin(kx'-\Omega t).
\eea   
Since the factor in parenthesis in the first line is seen
to be unity by use of the dispersion relation, we recover
the earlier expression for $\v_1$, and confirm 
that   the gauge invariant decomposition works as
advertised. Again we  see that the velocity field $\xi$,
which  arises from angular momentum conservation, is
smaller than the pressure induced flow, $\nabla\psi$, by a
factor of $\omega_0/\Omega$. 

\section{Discussion}

The central idea  in this paper is the decomposition
$\v_1=\nabla\psi+\xi $ of the velocity perturbation
into a potential flow  and a correction required by
angular momentum conservation.  This decomposition is
motivated by the Clebsch formalism, but does not depend on
it. From the decomposition  we see that corrections to the
acoustic metric  equation depend only on the ratio of the
frequency of the sound wave to the frequency, $\omega_0/2$,
of the background fluid rotation. This rotation frequency
is determined by the antisymmetric part, $(\partial_i v_j-
\partial_j v_i)/2 $, of the velocity inhomogeneity. The
symmetric part, the rate of strain  $e_{ij}= 
(\partial_i v_j+ \partial_j v_i)/2$,
can be large and the correction remain small. 
This is not  unreasonable  because
the acoustic metric  equation is exact for any potential
background flow --- no matter how inhomogeneous.  

At low frequencies  the correction $\xi= \x_1\times
\omega_0$ ceases to be  negligible. In this regime the
sound waves  hybridize with whichever of the many other
modes available to a fluid with vorticity happen to have
comparable frequency. The hybridization may lead to a
spectral gap, as with the Poincar{\'e} waves, to
birefringence, and to other phenomena which show that the
acoustic metric is no longer all that is needed to describe
sound propagation.

\section{Acknowledgements}

MS was supported by the NSF (USA) under grant DMR-98-17941.
SEPB was supported by  FAPERj (Brazil). MV was supported
by the US Department of Energy, and KEH by Ministerio de Educacion y Cultura,
Espa\~na. This work was
begun  during the  workshop {\it Analog models of General
Relativity\/} at CBPF,  the Brazilian Center for Research in
Physics, located in Urca, Rio de Janeiro, Brazil. MV and MS thank
the center for their hospitality and for hosting the
workshop.  


\end{document}